\documentclass[useAMS,usenatbib]{mn2e}
\usepackage{epsfig}

\title[FU Tau: Clues from variability]{Magnetic activity and accretion on FU Tau A: Clues from variability}

\author[Scholz et al.]{Alexander Scholz$^{1}$\thanks{E-mail: aleks@cp.dias.ie}, Beate Stelzer$^{2}$, Grainne
Costigan$^{1}$, David Barrado$^{3,4}$,  
\newauthor{Jochen Eisl{\"o}ffel$^{5}$, Jorge Lillo-Box$^{3}$, Pablo Riviere-Marichalar$^{3}$, Hristo Stoev$^{3}$}\\
$^{1}$ School of Cosmic Physics, Dublin Institute for Advanced Studies, 31 Fitzwilliam Place, Dublin 2, Ireland\\
$^{2}$ INAF - Osservatorio Astronomico di Palermo, Piazza del Parlamento 1, I-90134 Palermo, Italy\\
$^{3}$ Centro de Astrobiologia Depto. Astrofisica (INTA-CSIC), ESAC campus, P.O. Box 78, E-28691 Villanueva de la C\~anada, Spain\\
$^{4}$ Calar Alto Observatory, Centro Astron\'omico Hispano Alem\'an, Almer\'ia, Spain\\
$^{5}$ Th{\"u}ringer Landessternwarte Tautenburg, Sternwarte 5, D-07778 Tautenburg, Germany\\}

\begin{document}

\date{Accepted. Received.}

\pagerange{\pageref{firstpage}--\pageref{lastpage}} \pubyear{2002}

\maketitle

\label{firstpage}

\begin{abstract}
FU Tau A is a young very low mass object in the Taurus star forming region which was previously found
to have strong X-ray emission and to be anomalously bright for its spectral type. In this study we discuss 
these characteristics using new information from quasi-simultaneous photometric and spectroscopic monitoring. 
From photometric 
time series obtained with the 2.2\,m telescope on Calar Alto we measure a period of $\sim 4$\,d for FU Tau A, 
most likely the rotation period. The short-term variations over a few days are consistent with the rotational
modulation of the flux by cool, magnetically induced spots. In contrast, the photometric variability on 
timescales of weeks and years can only be explained by the presence of hot spots, presumably caused by 
accretion. The hot spot properties are thus variable on timescales exceeding the rotation period, maybe due 
to long-term changes in the accretion rate or geometry. The new constraints from the analysis of the variability 
confirm that FU Tau A is affected by magnetically induced spots and excess luminosity from accretion. However, 
accretion is not sufficient to explain its anomalous position in the HR diagram. In addition, suppressed 
convection due to magnetic activity and/or an early evolutionary stage need to be invoked to fully account for 
the observed properties. These factors cause considerable problems in estimating the mass of FU Tau A and other 
objects in this mass/age regime, to the extent that it appears questionable if it is feasible to derive the 
Initial Mass Function for young low-mass stars and brown dwarfs.

\end{abstract}


\begin{keywords}
stars: low-mass, brown dwarfs; stars: rotation; stars: activity; accretion, accretion discs
\end{keywords}

\section{Introduction}
\label{intro}

FU Tau has been known for several decades as a variable star embedded in the dark cloud B215 
\citep[e.g.][]{1961VeSon...5...87G,1975MmSAI..46...81R} and as a member of the Taurus star forming 
region based on H$\alpha$ emission and proper motion \citep{1953BOTT....1h...3H,1979AJ.....84.1872J}. 
Recently the object has drawn interest because it turned out to be a binary, with a primary component 
FU Tau A and the faint companion FU Tau B at a separation of 5.7", corresponding to a projected 
separation of $\sim 800$\,AU for the distance of Taurus. \citet{2009ApJ...691.1265L} discovered the companion 
and estimated spectral types of M7.25 and M9.25 and masses of 0.05 and 0.015$\,M_{\odot}$ for the two 
objects, indicating that FU Tau may in fact be a rare wide binary brown dwarf. Because of the wide 
separation of the two components and the location far away from any other member of the star 
forming region, FU Tau is of considerable interest to test formation scenarios for substellar 
objects.

Apart from its binarity and location, FU Tau A turns out to exhibit two other anomalous properties.
First, the object has strong X-ray emission compared with other objects at similar spectral type 
\citep{2010MNRAS.408.1095S}. Moreover, the X-ray spectrum indicates the presence of soft radiation, 
possibly from an accretion-related shockfront, as has been observed previously for more massive 
objects \citep[e.g.][]{2004A&A...418..687S}. Second, FU Tau A is anomalously bright for objects of 
this spectral type in the Taurus star forming region. It sits about one order of magnitude above
the 1\,Myr isochrone in the HR diagram \citep{2009ApJ...691.1265L,2010MNRAS.408.1095S}. In 
\citet{2010MNRAS.408.1095S} a few possible scenarios have been put forward to explain these features, 
including suppressed convection due to magnetic activity, excess flux from accretion, and early
evolutionary stage.

Here we set out to put further constraints on the properties of FU Tau A by analysing its photometric
and spectroscopic variability. This paper is mainly based on photometric time series obtained with 
the instruments CAFOS and BUSCA at the 2.2\,m telescopes of the German-Spanish Astronomical Center
at Calar Alto observatory. In Sect. \ref{data} we discuss these observations and the reduction of the data. 
The analysis of the photometry and spectroscopy is presented in Sect. \ref{phot} and \ref{spec}. In Sect.
\ref{disc} we compile all available information on the variability of the system from our new observations,
the literature, and archives and constrain the origin of the variations using spot models. We
discuss the results in the context of the two anomalies mentioned above in the final 
Sect. \ref{nature}.

\section{Observations and data reduction}
\label{data}

\subsection{CAFOS}

\subsubsection{Imaging}

Our primary photometric time series was obtained with CAFOS at the 2.2\,m telescope on 
Calar Alto over five nights in Nov/Dec 2010. CAFOS is a $2\times 2$k CCD camera mounted 
in the RC focus. With a pixel scale of 0\farcs53, it gives a field of view (FOV) of 
$16'\times 16'$. The filters, however, do not cover the full FOV; in effect a circular 
FOV with diameter of $\sim 14'$ can be used.

While the whole run was affected by dodgy weather conditions,
including bad seeing, high humidity, and clouds, we obtained 62 useful images in
the R-band and the same number in the I-band for our target. The final night of
the run we observed Landolt standard stars under photometric conditions for calibraton
purposes (2x field SA92, 3x field SA98). The observing log for the run is given in 
Table \ref{obs}.

FU Tau is located in the middle of a dark cloud devoid of stars. Since we need non-variable
field stars to calibrate the lightcurves, our FOV was not centered on FU Tau itself. 
Instead, we positioned FU Tau in the south-west (SW) corner of the CCD, which allows us to 
cover a sufficient number of field stars in the area immediately north-east (NE) of the cloud.
To minimize flatfield problems, we aimed to keep the position of the time series field as 
constant as possible; the offsets between the images are $<10"$. We also aimed to keep the 
flux level of FU Tau A roughly constant, i.e. we varied the exposure times to account for 
changes in seeing and transparency.

For all images we carried out a standard reduction: subtracting the average bias
level and dividing by a scaled, averaged domeflat. We found that the flatfielding is
not perfect; the resulting frames are affected by a large-scale interference pattern,
which might be due to water condensation on the detector window. The effect increases
with the time offset between science frames and domeflats. Therefore we obtained two
sets of domeflats per night and corrected the images using the flatfield with the
minimum time offset. Since the pattern has a spatial scale of $>50"$ and the spatial
offsets in the time series are $<10"$ we do not think that the pattern has
an effect on the lightcurves. In addition, the I-band frames show a faint small-scale
interference pattern due to nightsky emission lines, which contributes to the noise.

\subsubsection{Spectroscopy}

As part of the same run, we obtained 5 low-resolution spectra for FU Tau A, using
grism R400 with a nominal resolution of 10\,\AA. 

The spectra for FU Tau A were debiased and background-subtracted by fitting a 2nd order
polynomial to each line in the spatial direction. They were extracted, dispersion-corrected
and flux-calibrated using standard routines in IRAF. 

\begin{table}
\caption{Time series observations with CAFOS and BUSCA. In the 2nd column, the 'C' stands
for CAFOS and the 'B' for BUSCA. The values for exposure times and seeing are typical
for a particular night.
\label{obs}}
\begin{tabular}{ccccc}
\hline
Date         & bands  & no. & exp time  & seeing \\
\hline
2010-11-28   & C/R,I  & 6   &  250, 100 & 2"   \\
2010-11-30   & C/R,I  & 21  &  450, 120 & 3-4" \\
2010-12-01   & C/R,I  & 8   &  300, 100 & 2"   \\
2010-12-02   & C/R,I  & 27  &  300,  80 & 2"   \\
\hline
2010-12-08   & B/I    & 2   &  120      & 2\farcs1 \\
2010-12-10   & B/I    & 5   &  120      & 1\farcs7 \\
2010-12-11   & B/I    & 6   &  120      & 1\farcs6 \\ 
2010-12-13   & B/I    & 1   &  120      & 1\farcs5 \\ 
2010-12-15   & B/I    & 4   &  120      & 2\farcs4 \\ 
\hline
\end{tabular}
\end{table}

\subsection{BUSCA}

Complementary time series photometry in the I-band was obtained using BUSCA at the 2.2\,m telescope
on Calar Alto. BUSCA allows to take images in four bands simultaneously, achieved through 3
dichroic beam splitters. Our target, however, was not detected in the three blue channels (Stromgren
vby filters); we only use the images in the reddest channel, which corresponds to the Cousins I-band. 
The observations started about a week after the CAFOS run and continued for another week. Similar to 
the CAFOS run, parts of the observations were affected by clouds, strong winds, and high humidity. No 
photometric calibration was carried out.

An $11'\times 11'$ FOV centered on FU Tau was observed in 7 nights in Dec 2010, of which 5 provided
usable data (see Table \ref{obs}). The FOV covers the bright K2III star 2MASS J04232455+2500084 and 
5-10 point sources 1-2\,mag fainter than FU Tau A. For all images, we carried out a standard reduction 
including bias subtraction and flatfield correction. 

\section{Time series photometry}
\label{phot}

\subsection{Photometry and relative calibration}
\label{relcal}

From the CAFOS time series we derived R- and I-band lightcurves for FU Tau A. We hand-picked a sample
of 48 (I) and 45 (R) sources, including FU Tau A and all other isolated stars with similar brightness 
in the FOV. For these objects we carried out aperture photometry using a constant aperture of 10\,pixel
and a sky annulus of 10-20\,pixel. Due to the poor seeing, the companion FU Tau B is not detected in most 
of the CAFOS images; no photometry was possible for this object. 

\begin{figure*}
\includegraphics[width=6.1cm,angle=-90]{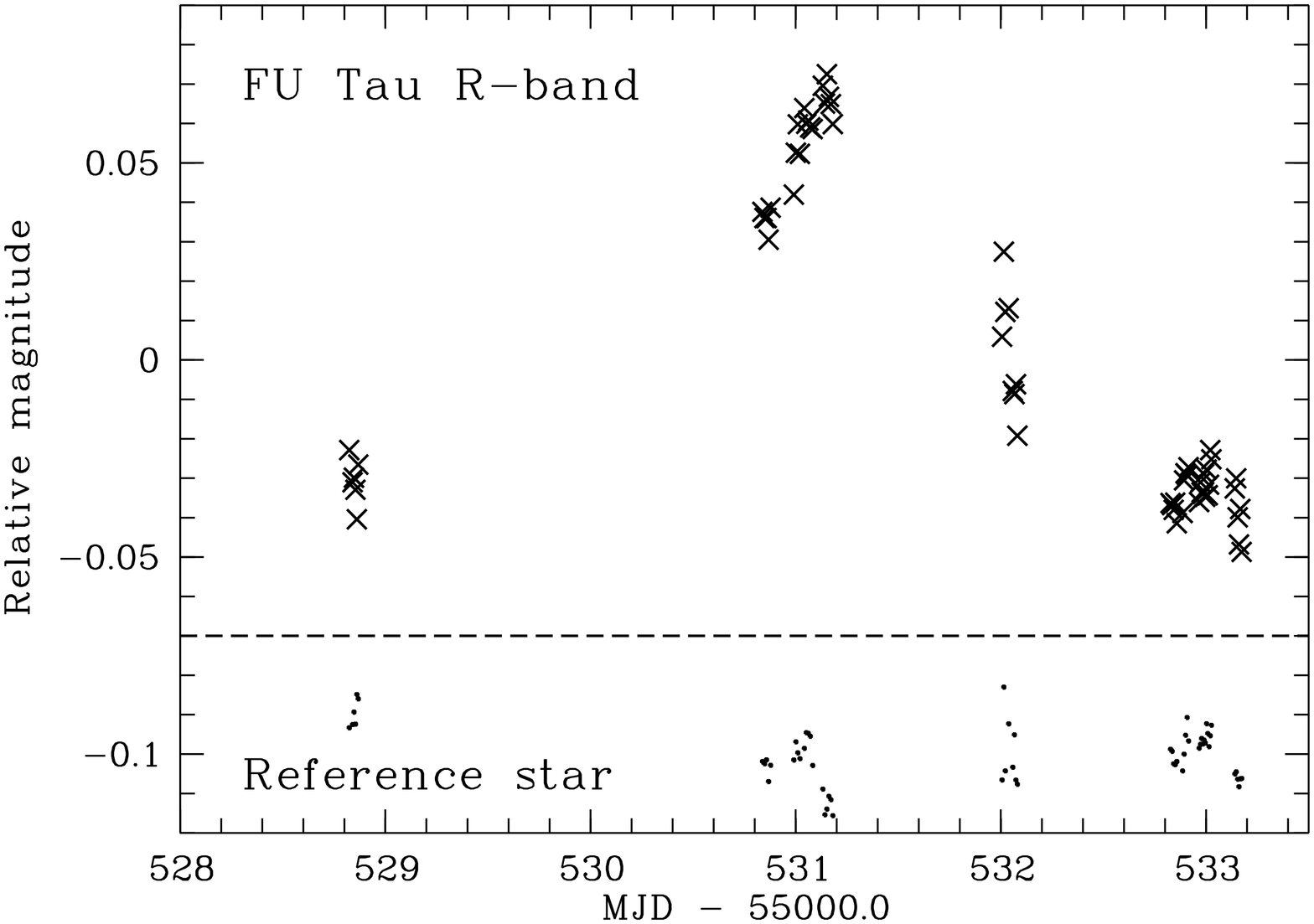} \hfill
\includegraphics[width=6.1cm,angle=-90]{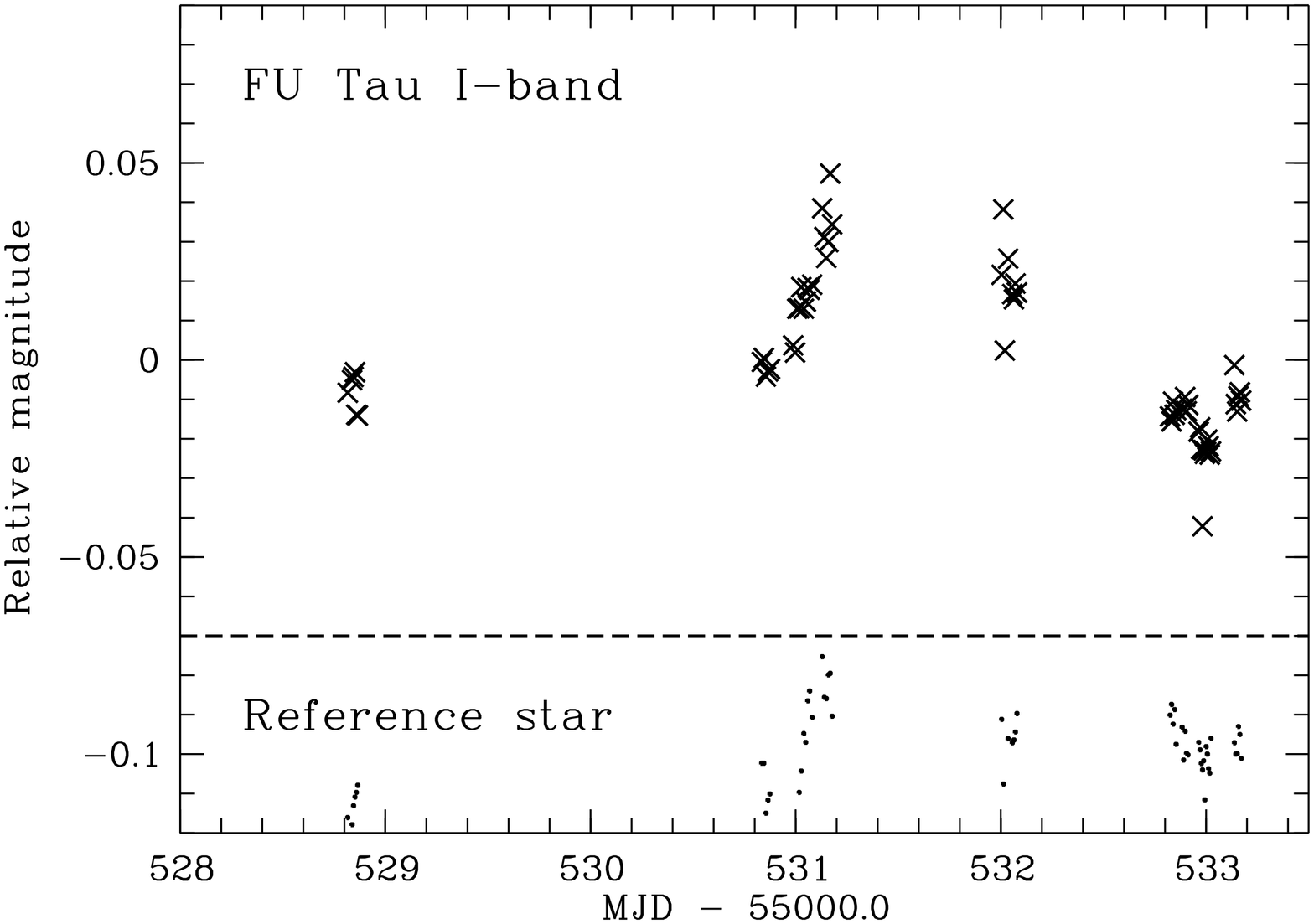} 
\caption{Lightcurves for FU Tau A derived from the CAFOS images for the R- and the I-band. In addition
to the target, we show the lightcurve of a similarly bright, non-variable reference star in the same
field. The intra-night variability of FU Tau A is partly seen in the reference stars as well and could be
due to secondary extinction effects and not intrinsic to the source. The inter-night variability, 
however, is not apparent in the reference stars.
\label{f1}}
\end{figure*}

To correct for the effects of variable seeing and transparency ('relative calibration'), we calculated 
the average time series of non-variable stars in the field and subtracted it from all lightcurves. The 
non-variable stars were chosen using the procedure outlined in \citet{2004A&A...419..249S}. The routine 
selected 18 (R) and 10 (I) stars as non-variable, based on a comparison of their lightcurve with the 
average lightcurve of all other stars. The average RMS of the lightcurves for these non-variable stars 
is 0.010 (R) and 0.012\,mag (I), which defines the photometric accuracy. 

\begin{figure}
\includegraphics[width=6.1cm,angle=-90]{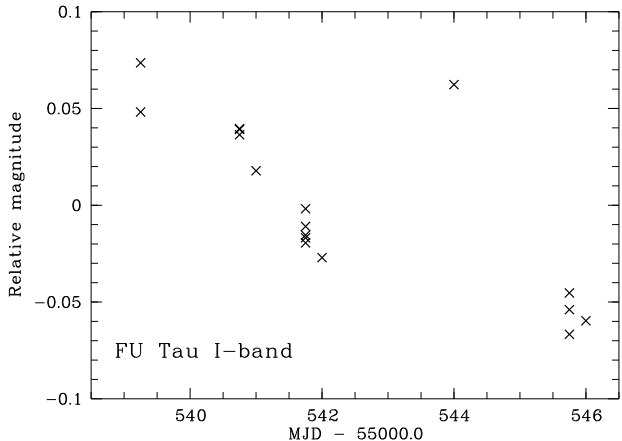} \hfill
\caption{I-band lightcurve for FU Tau A derived from the BUSCA images. 
\label{f4}}
\end{figure}

From the BUSCA images we obtained I-band lightcurves for FU Tau A and all other stars in the field, again 
using aperture photometry with the same parameters as for CAFOS. The bright star 2MASS J04232455+2500084
clearly looks variable, but 7 faint field stars show stable lightcurves. Their average lightcurve is used
for the relative calibration. After subtraction of the average lightcurve, the RMS for the 7 
field stars is 0.011-0.025, an average of 0.018\,mag, confirming that they are non-variable. For 
comparison, the RMS for 2MASS J04232455+2500084 is 0.13\,mag.

\subsection{Lightcurve analysis}
\label{lc}

The lightcurves from CAFOS and BUSCA show that FU Tau A is a variable star. Its RMS is 0.04 (R-band,
CAFOS), 0.02 (CAFOS, I-band), and 0.04 (BUSCA, I-band), significantly more than comparison stars  
(0.01\,mag for CAFOS, 0.02\,mag for BUSCA, Sect. \ref{relcal}). The lightcurves from CAFOS and BUSCA are
shown in Figs. \ref{f1} and \ref{f4}. Most of the variability is on timescales of 
$>1$\,d; these variations are intrinsic to the source and are not seen in the reference stars. 

In addition, the CAFOS lightcurves show intra-night variability with smaller amplitude. These variations, 
however, are partly seen in the reference stars as well. FU Tau A is by far the reddest object in our 
sample. The influence of the atmospheric conditions is colour-dependent, which is not taken into account 
in our correction. Thus, one could expect the reddest objects to show some residuals of the trends
caused by atmospheric effects. Hence, we do not consider the low-level intra-night variability in FU 
Tau A to be real.

\begin{figure*}
\includegraphics[width=6.1cm,angle=-90]{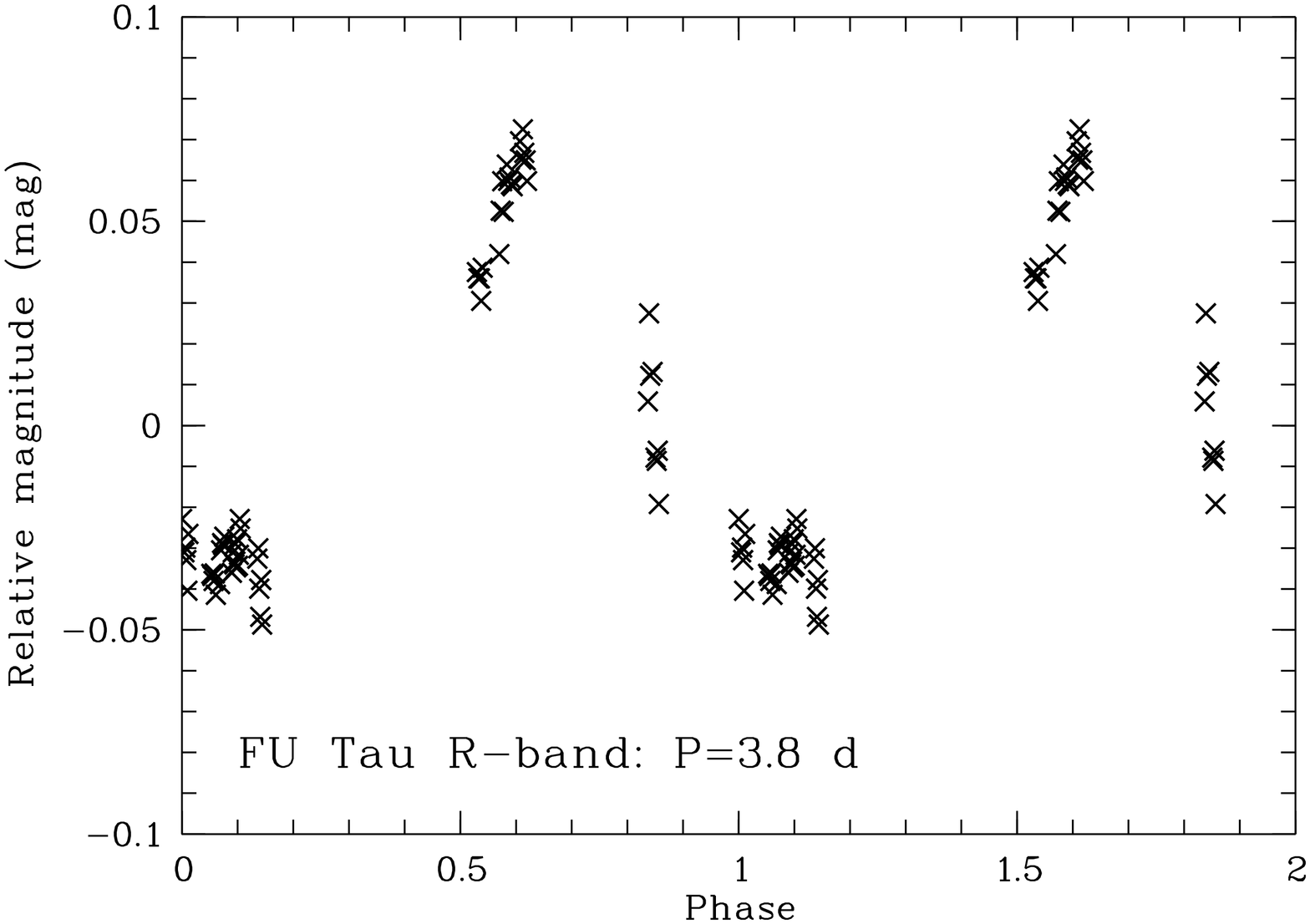} \hfill
\includegraphics[width=6.1cm,angle=-90]{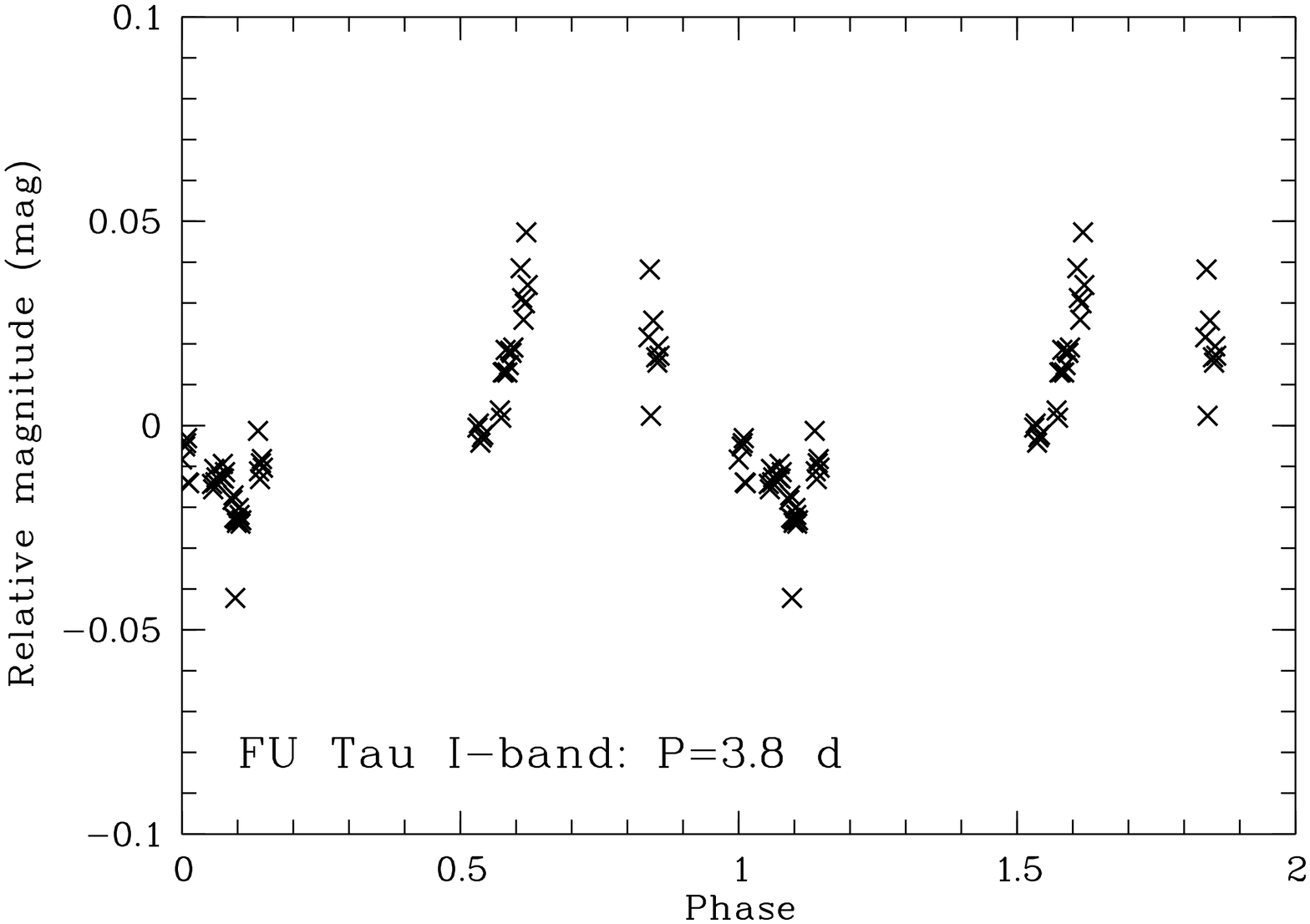} \hfill
\caption{Phase-folded lightcurves for FU Tau A assuming our best-fitting period of 3.8\,d, determined
from the CLEANed periodograms. The typical error is 0.010\,mag in R and 0.012\,mag in I.
\label{f3}}
\end{figure*}

Using all available CAFOS datapoints for a given filter, we searched for a period using a combination of 
three routines. The R- and the I-band lightcurves show a dominant peak in the CLEANed periodogram 
\citep{1987AJ.....93..968R} at a period of 3.8 (R) and 4.0\,d (I). The same peak is detected in the 
Scargle periodogram \citep{1982ApJ...263..835S} with a false alarm probability below $10^{-5}$ 
(calculated following \citet{1986ApJ...302..757H}). In the Scargle periodogram, however, the peak is 
very broad and does not permit an accurate assessment of the period. Finally, we compare the RMS in the 
original lightcurve with the RMS after subtracting a sine function with the suspected period using the 
F-test. Again, the period of 3.7-4.0\,d is highly significant in both bands, with false alarm probabilities 
below $10^{-5}$. In Fig. \ref{f3} we show the phase-folded lightcurve assuming $P=3.8$\,d, which we 
consider to be the best-fitting period from all three algorithms. The observing run covers 
only one period, and the sampling of the period is patchy (in phase space). Therefore, a relatively large 
range of periods (3.5-4.1\,d) give a decent fit to the data, i.e. the uncertainty in the period 
is in the range of $\pm 0.3$\,d.

\begin{figure*}
\includegraphics[width=6.1cm,angle=-90]{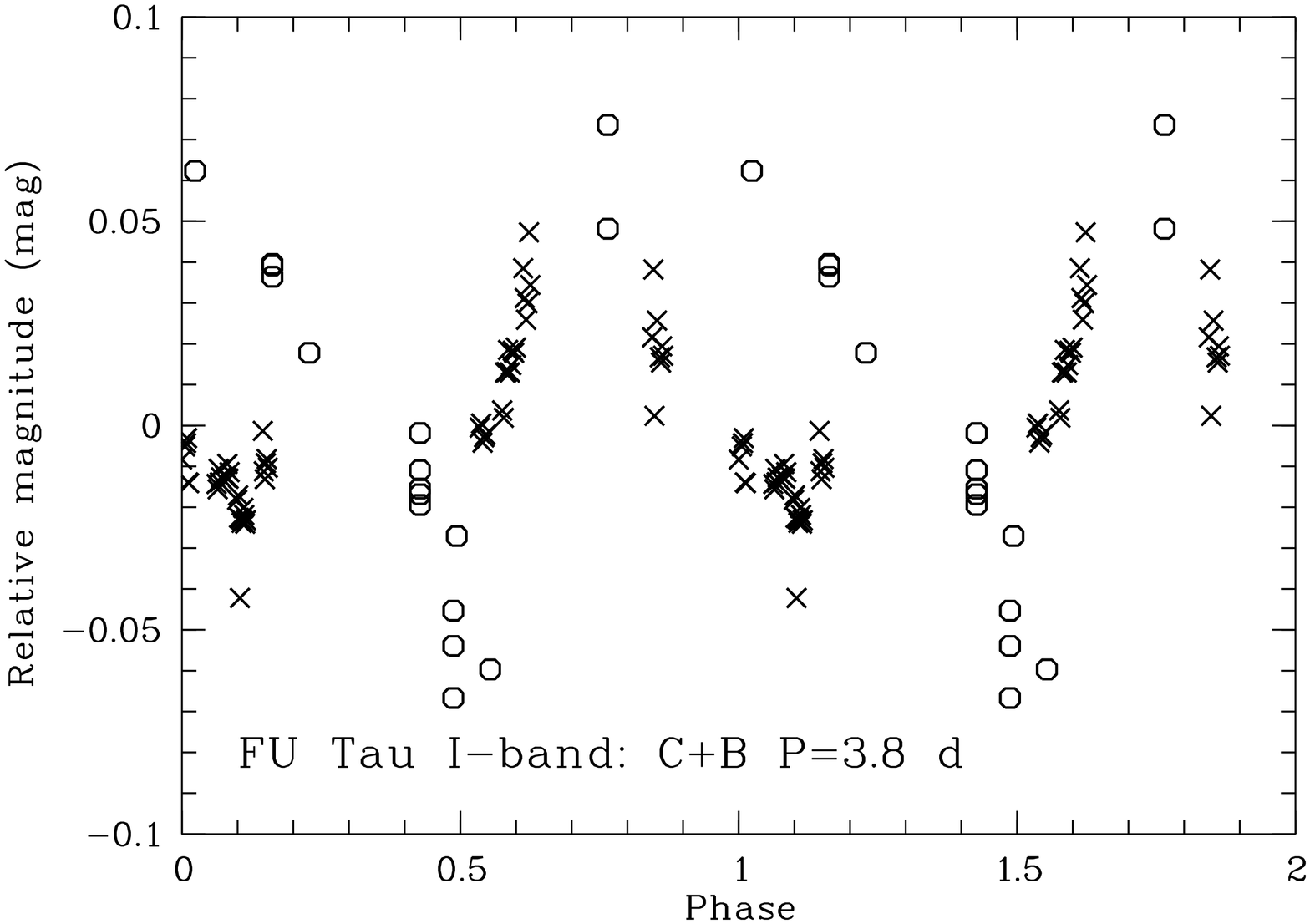} \hfill
\includegraphics[width=6.1cm,angle=-90]{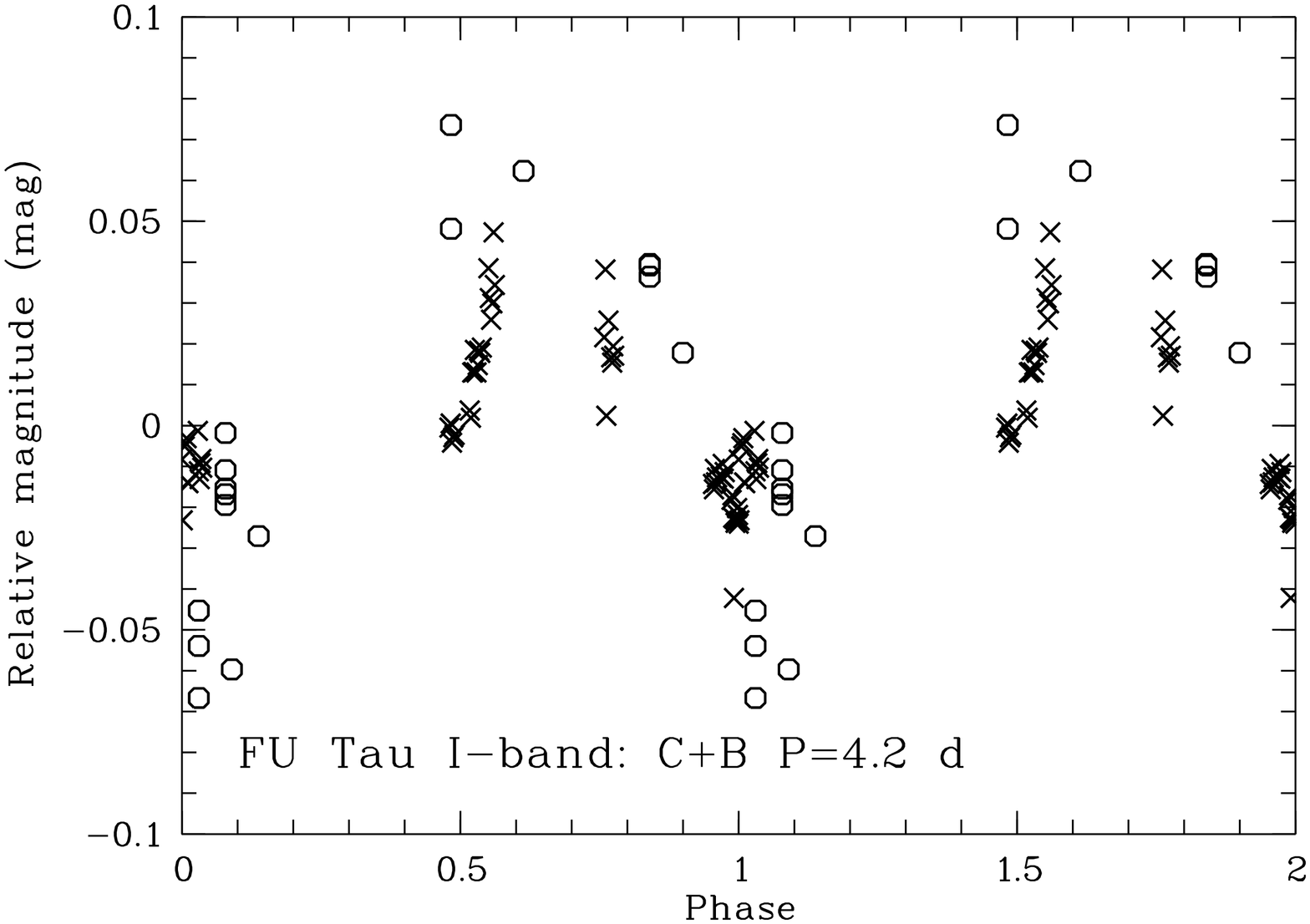} \hfill
\caption{Phase-folded I-band lightcurves for FU Tau A showing the datapoints from CAFOS (crosses) and BUSCA
(circles) for two different periods. The typical error is 0.012\,mag for CAFOS and 0.018\,mag for BUSCA.
\label{f5}}
\end{figure*}

Although the coverage with BUSCA is not sufficient to carry out an independent period search, we use
these datapoints to check the period derived from the CAFOS lightcurves. In Fig. \ref{f5} we show a phaseplot
for all I-band data, assuming a period of 3.8\,d (left panel) and 4.2\,d (right panel).  A good match is 
achieved for a period of 4.2\,d, slightly larger than the period determined from the R- and I-band CAFOS 
lightcurves. A period of 3.8\,d only matches if a significant phase shift between the CAFOS and the BUSCA
data is assumed.

\subsection{Calibrated photometry}
\label{calphot}

From the Landolt standard fields observed in the last night of the CAFOS run we derived a photometric
calibration. In total, we observed 45 standards from which 40 gave useful photometry. These stars cover
a wide range in airmass from $X=1.3$ to 2.0. For the R-band the absolute magnitudes $R$ are well reproduced with
a zeropoint shift and an extinction term: $R = r - 1.647 - 0.147X$. The RMS for this transformation
is 0.03, dominated by the uncertainty in the zeroterm. For the I-band, it turns out that an additional
colour term improves the RMS from 0.1 to 0.05: $I = i - 2.281 - 0.069X + 0.136(r-i)$. (In these
equations, the lower case letters are instrumental magnitudes and upper case letters calibrated
magnitudes.)

Applying this transformation to the instrumental magnitudes measured for FU Tau A gives $R = 15.39$ 
and $I = 13.73$\,mag for the night 2010-12-02. For this night the lightcurve for FU Tau A indicates
a photometric uncertainty of $\sim 0.02$\,mag (see Sect. \ref{lc}). Adding this in quadrature to
the calibration errors, the total uncertainty in the absolute magnitudes is 0.04 in the R-band and 0.05 
in the I-band.

Published photometry in similar bands for FU Tau A is available from CFHT (Cousins I), Sloan (r, i),
and the Carlsberg Meridian Catalog 14 (filter close to Sloan r). To transform the Sloan magnitudes 
to the Johnson/Cousins system, we used Equ. (2) and (8) from \citet{2006A&A...460..339J}. All 
calibrated photometry in the bands R and I is listed in Table \ref{cal}. The band transformations 
from Sloan to Cousins depend linearly on $R-I$ and are only calibrated for $0<R-I<2$, whereas FU Tau 
A is slightly redder ($R-I=2.3-2.5$) in the Sloan photometry. There is, however, no evidence for 
additional colour trends for $R-I<2$ in Fig. 3 of \citet{2006A&A...460..339J}, i.e. any error 
introduced by this conversion is likely to be small.

\begin{table}
\caption{Calibrated photometry for FU Tau A in bands similar to Johnson R and I. Typical
errors are 0.05\,mag. \label{cal}}
\begin{tabular}{llllll}
\hline
Date             & Sloan r & Sloan i& John R & John I& Comment\\
\hline
Oct 2001         & 16.94   &        &       &        & CMC14$^1$\\
06/12/2002       & 17.13   & 14.86  & 16.41 & 13.92  & Sloan$^{2,3}$\\
29/12/2002$^4$   & 16.86   & 14.75  & 16.18 & 13.85  & Sloan$^{2,3}$\\
29/12/2002       &         &	    &	    & 13.58  & CFHT$^2$\\
02/12/2010       &         &	    & 15.39 & 13.73  & CAFOS$^5$\\
\hline
\end{tabular}

$^1$ \citet{2002A&A...395..347E}\\
$^2$ \citet{2009ApJ...691.1265L}\\
$^3$ Sloan magnitudes converted to Johnson using equations in \citet{2006A&A...460..339J}\\
$^4$ Conflicting epoch information in \citet{2009ApJ...691.1265L} (29 or 31/12/2002)\\
$^5$ this paper\\
\end{table}

The two epochs of Sloan photometry differ by 0.07 in $I$ and 0.23 in $R$ over a timescale of 23-25\,d, which 
is slightly more than our amplitudes measured over 5 nights. Comparing our photometry with the Sloan values 
indicates long-term variability of 0.2\,mag in $I$ and 0.9\,mag in $R$. FU Tau A was 
much fainter and also redder in $R-I$ in 2002 compared with 2010 ($R-I\sim 2.4$ vs. 1.7\,mag). 

The Sloan and CFHT photometry has been measured in December 2002. In \citet{2009ApJ...691.1265L} 
the epochs are listed as Dec 6 and 29 in their Table 1 and as Dec 6 and 31 in the text. 
FU Tau is not listed in the 6th data release of the Sloan Digital Sky Survey \citep{2008ApJS..175..297A}
referenced by \citet{2009ApJ...691.1265L}, and not in the 7th data release either, therefore
we cannot verify the exact source of the photometry. According to \citet{2009ApJ...691.1265L} the 
CFHT data is from Dec 29.

The I-band difference between the second Sloan epoch and CFHT of 0.27\,mag seems quite a lot compared 
with the low level of variability in our lightcurve ($\le 0.05$\,mag in one night, 0.09\,mag over 4 nights, 
Sect. \ref{lc}). The large difference could be due to a) inconsistencies in the I-band calibration, b) 
problems with the transformation from Sloan to Cousins bands (see above), or c) a strong outburst in 
that particular night. For these reasons we cannot reliably use the CFHT I-band magnitude and disregard 
the datapoint in the following. The CMC14 datapoint was obtained in October 2001 and is consistent with 
the Sloan r-band value from December 2002.

\section{Spectroscopy}
\label{spec}

\begin{figure}
\includegraphics[width=6.1cm,angle=-90]{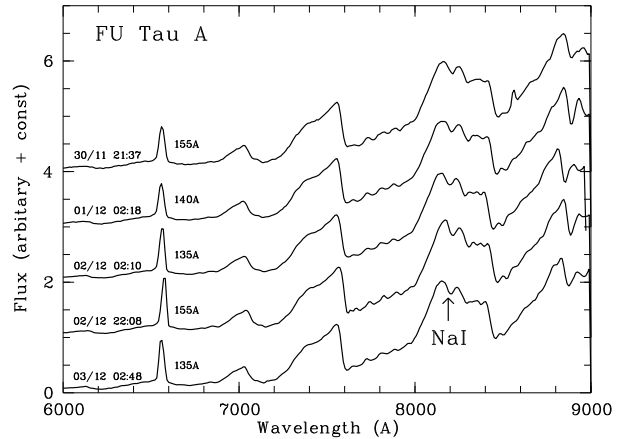} \hfill
\caption{Low-resolution spectroscopy for FU Tau A from CAFOS. The dates and times of the observations
and the equivalent widths for H$\alpha$ are indicated.  The NaI absorption feature is marked. The 
resolution is 10\,\AA.
\label{f2}}
\end{figure}

The five CAFOS spectra for FU Tau A, obtained in 4 different nights in Nov/Dec 2010, are shown in Fig. 
\ref{f2}. The spectral shape is remarkably similar; we do not find any significant differences in the 
continuum. With our low resolution of 10\,\AA, the H$\alpha$ feature at 6563\,\AA\,is the only emission 
feature. The equivalent widths (EW) for this line are in the range of 135-155\,\AA\,and 
are indicated in Fig. \ref{f2}. The variations in the EW are not significantly larger than the uncertainty, 
which is in the range of $\pm 10-20$\,\AA. The EWs are also consistent with the value of 146\AA\,measured 
from a Gemini/GMOS spectrum in March 2008 \citep{2010MNRAS.408.1095S}. 

Using the PC3 index suggested by \citet{1999AJ....118.2466M} we assigned spectral types. In chronological
order, the spectral types are M6.82, M6.86, M6.67, M6.72, M6.66, i.e. the change is only $\pm 0.1$ subtypes. 
Compared with the uncertainties, this is not a significant variation. Thus, the photometric variability
observed for FU Tau A does not manifest itself as a change in spectral type. This is not particularly 
surprising, as the colour variation in $R-I$ over this time series is only 3\% (Table \ref{amps}).

The H$\alpha$ EW of FU Tau A is well-above the usually adopted threshold between non-accreting 'weak-line' and
accreting 'classical' T Tauri stars (CTTS). According to \citet{2003AJ....126.2997B}, all M7 objects with H$\alpha$
EW $>40$\,\AA~should be considered to be accretors. FU Tau A's status as substellar analog to a CTTS is confirmed 
by the presence of mid-infrared excess \citep{2009ApJ...691.1265L}, most likely from a circumstellar disk, and 
a wealth of other accretion-related emission lines in the optical and near-infrared spectrum (Stelzer et al., 
in prep.). 

The spectra also clearly show the NaI absorption features at $\sim 8200$\,\AA, which is in fact a doublet of 
lines at 8183 and 8195\,\AA \citep{1991ApJS...77..417K}. The feature is sensitive to surface gravity and can be 
used as an age indicator for cool stars. In our spectra the EW for NaI are 3-3.5\,\AA. For comparison, objects 
with similar spectral type as FU Tau A have typical NaI EW of 6-8\,\AA~in the field, 5-6\,\AA~in the 100\,Myr old 
cluster Pleiades, and 1.5-3.5 in the 3\,Myr old $\sigma$\,Ori cluster
\citep{1995MNRAS.272..630S,2005MNRAS.356...89K}. Thus, our NaI measurement confirms the youth of FU Tau A.

\section{Origin of the variability}
\label{disc}

The available photometry for FU Tau A provides an account of its variability on timescales ranging
from hours to years. While our photometric time series with CAFOS and BUSCA covers the short-term
variations (a week or less) in the optical, the archived data in the literature from Sloan and Spitzer 
constrains the long-term changes in the optical and infrared. In Table \ref{amps} we compile the variability 
amplitudes for FU Tau A from a variety of sources. In the following we aim to use the characteristics of 
this dataset to constrain the physical properties of FU Tau A, in particular the presence of cool spots 
caused by magnetic activity and/or hot spots caused by the accretion flow.

\begin{table}
\caption{Photometric amplitudes for FU Tau A as a function of filter and timescale
\label{amps}}
\begin{tabular}{llll}
\hline
Filter         & $\lambda$ ($\mu m$) & $\Delta$t & A (mag)		     \\
\hline
Johnson I      &  0.85               & 4\,d	 & $0.09\pm0.02$$^1$	     \\
Johnson R      &  0.65               & 4\,d	 & $0.12\pm0.02$$^1$	     \\
Johnson I      &  0.85               & 7\,d	 & $0.14\pm0.03$$^1$	     \\
\hline
Sloan z        &  0.89               & 23\,d	 & $0.07 \pm 0.07$$^2$       \\
Sloan i        &  0.75               & 23\,d	 & $0.09 \pm 0.07$$^{2,3}$   \\
Sloan r        &  0.62               & 23\,d	 & $0.27 \pm 0.07$$^{2,4}$   \\
Sloan g        &  0.47               & 23\,d	 & $0.44 \pm 0.07$$^2$       \\
Sloan u        &  0.36               & 23\,d	 & $0.71 \pm 0.08$$^2$       \\
\hline         
Johnson I      &  0.85               & 8\,yr	 & $0.2 \pm 0.1$$^{1,2}$     \\
Johnson R      &  0.65               & 8\,yr	 & $0.9 \pm 0.1$$^{1,2}$     \\
\hline
IRAC2          &  4.5                & 2\,yr	 & $0.19 \pm 0.03$$^2$       \\
IRAC4          &  8.0                & 2\,yr	 & $0.25 \pm 0.04$$^2$       \\ 	
\hline
\end{tabular}

$^1$ this paper\\
$^2$ \citet{2009ApJ...691.1265L}\\
$^3$ translates to Johnson I amplitude of 0.07\\
$^4$ translates to Johnson R amplitude of 0.23\\
\end{table}

As demonstrated in this paper, FU Tau A shows small-scale variations of $\sim 0.1$\,mag in the I- and R-band on 
timescales of a few days, with a periodicity in the range of 4\,d (Sect. \ref{lc}). A plausible explanation 
for this behaviour is the presence of asymmetrically distributed surface spots which cause a modulation of the
observed flux due to the rotation of the object, a behaviour typical for many young low-mass stars 
\citep{2007prpl.conf..297H}. The amplitudes are slightly larger in the R-band by factor of 1.3, as expected 
for cool spots \citep{1995A&A...299...89B}. 

Simple spot simulations were used to constrain the properties of the spots. We calculate the flux ratio between the
spotless surface of FU Tau and the surface with a single spot with a certain temperature $T_S$ and filling factor 
$f$ (defined as the fraction of the surface covered by the spot). For the spectrum of the unspotted photosphere, we 
used the AMES-DUSTY spectrum \citep{2001ApJ...556..357A} for $T=3000\,K$ and $\log{g} = 3.5$, typical of very young 
objects. Note that the actual temperature of the unspotted photosphere is not known. Based on the average spectral type 
of M6.75 (Sect. \ref{spec}), we would estimate 2800\,K \citep{2008ApJ...689.1127M}, but if spots are present the unspotted 
photosphere will be hotter. We ran the same simulations for $T=2800$ and 3200\,K. Since this gives only marginally 
different results, the exact choice of $T$ does not seem to be relevant.

\begin{figure*}
\includegraphics[width=6.1cm,angle=-90]{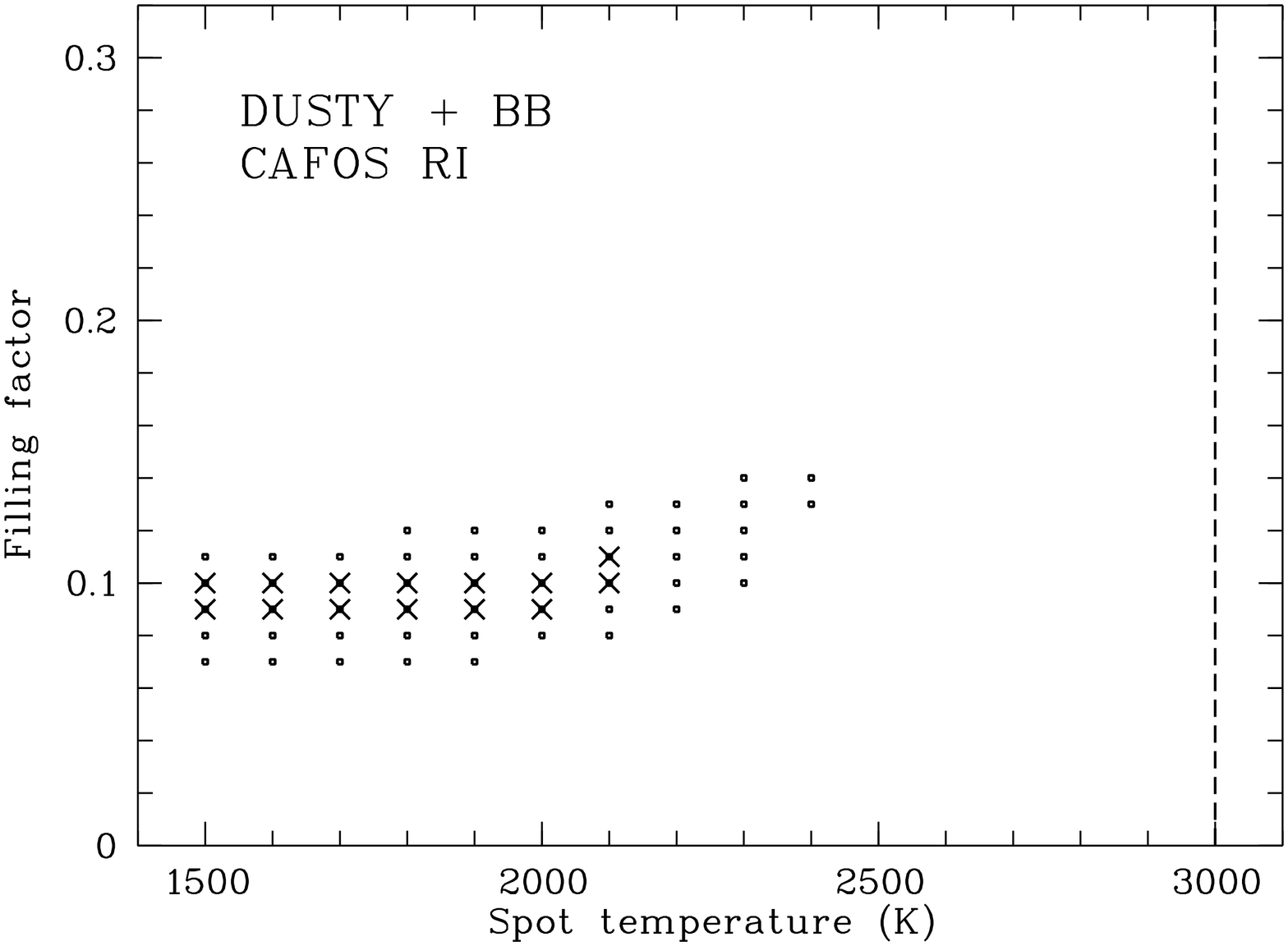} \hfill
\includegraphics[width=6.1cm,angle=-90]{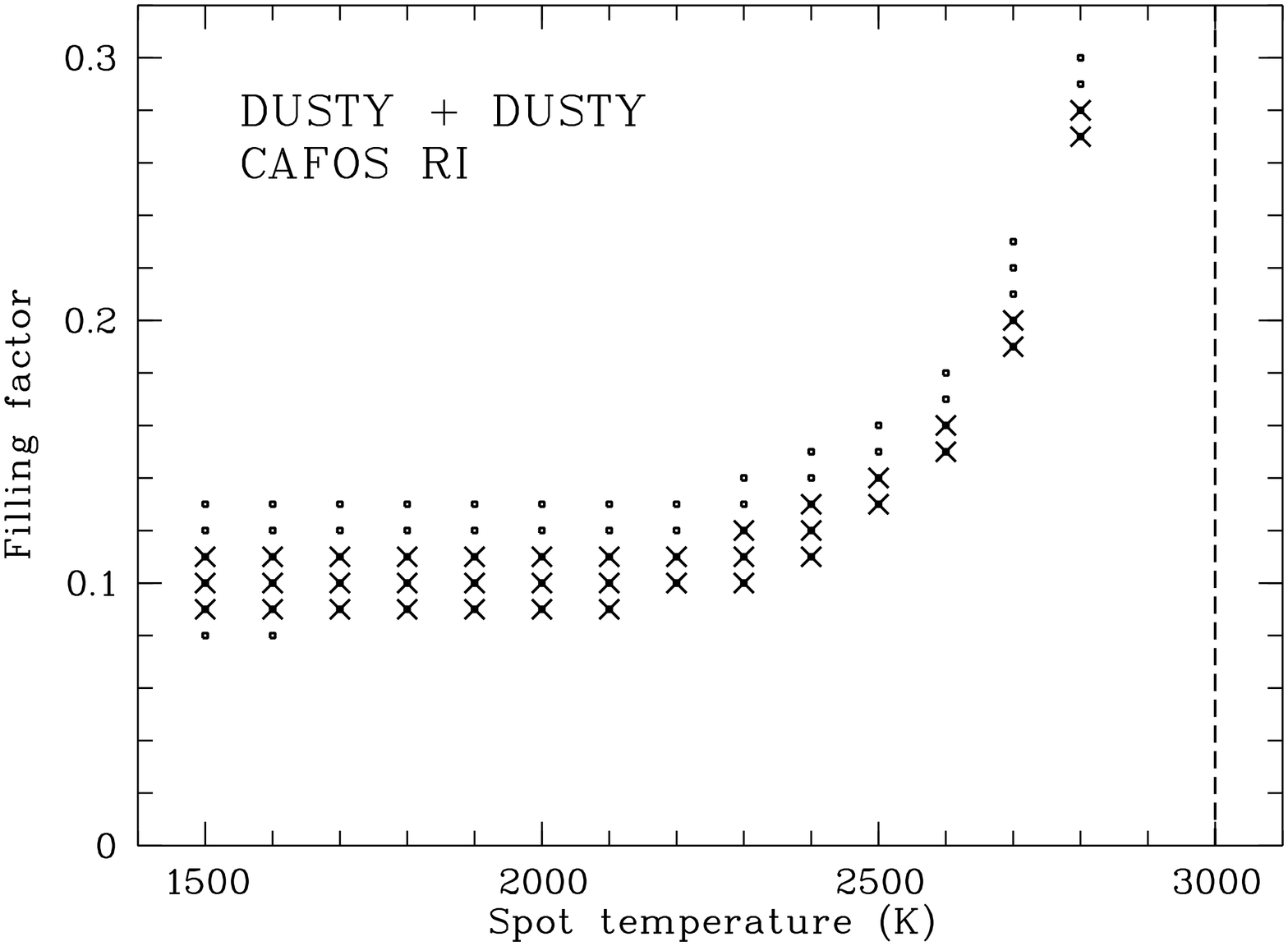} \\
\includegraphics[width=6.1cm,angle=-90]{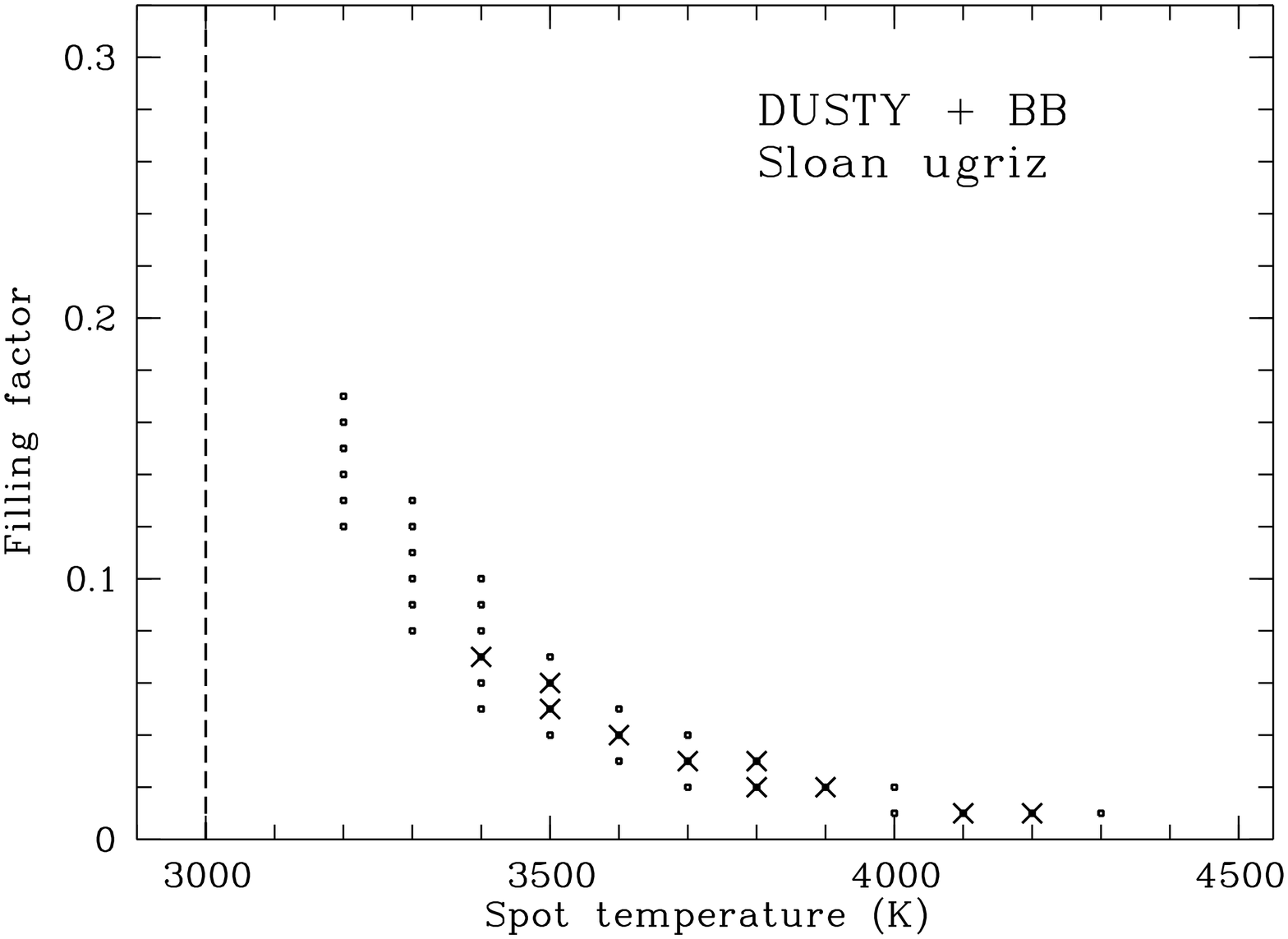} \hfill
\includegraphics[width=6.1cm,angle=-90]{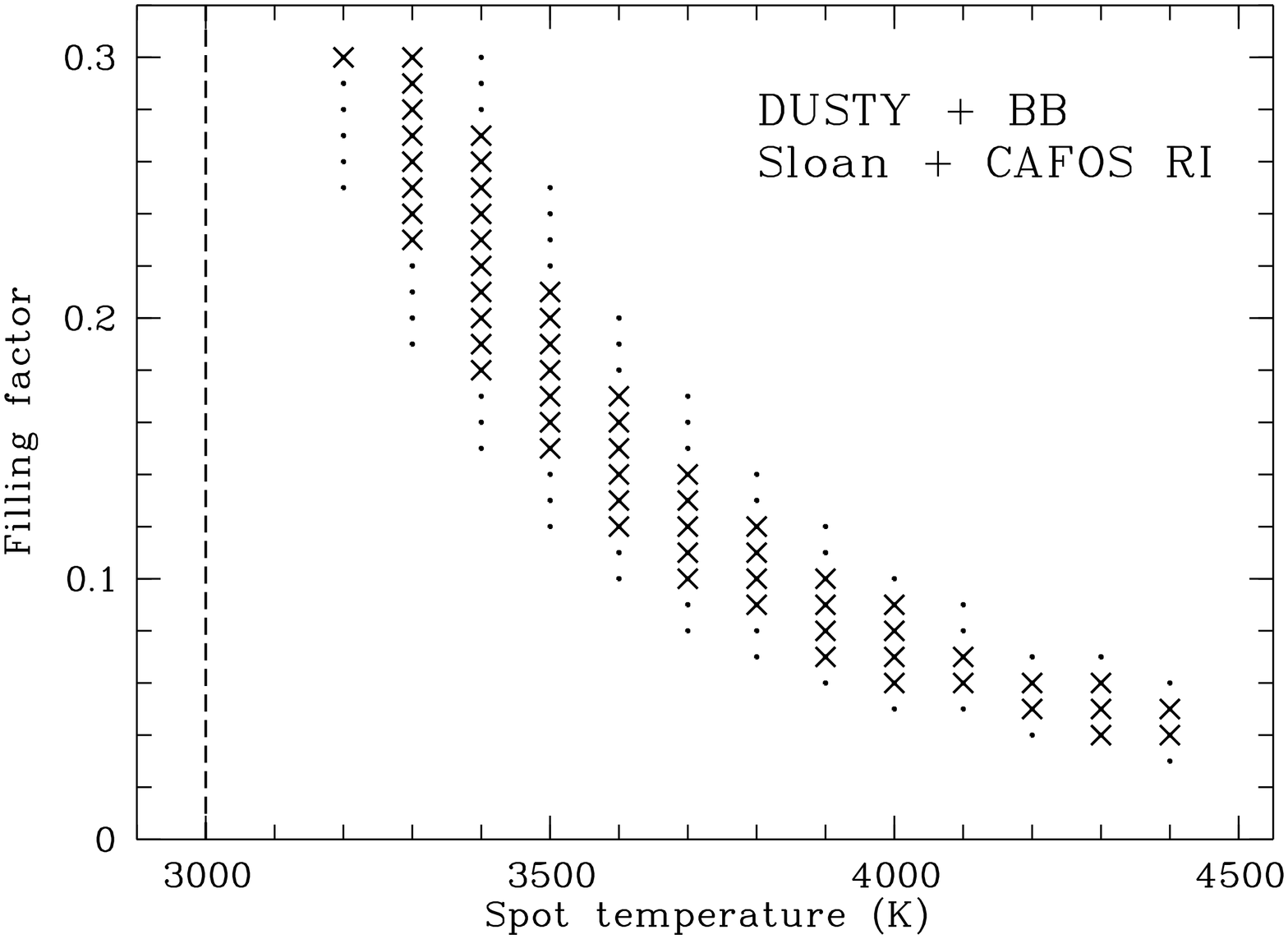} \\
\caption{Best-fitting spot temperature/filling factor combinations from the simulations discussed
in Sect. \ref{disc}. All crosses show the best fitting combinations with $\chi^2<1$, all dots with 
$\chi^2<3$.
\label{f6}}
\end{figure*}

The spot spectrum was approximated either by the AMES-DUSTY spectrum or with a blackbody. We calculated 
flux ratios as a function of wavelength for $T_S$ ranging from 1500 to 4000\,K in steps of 100\,K and $f$ ranging 
from 0.01 to 0.3 in steps of 0.01. For the blackbody spot, we extended the temperature range to 4500\,K.

To compare these ratios with the observations, we calculated the amplitudes $m_R$ and $m_I$ for the wavelengths
of the respective filters and derived the following test quantity: 
\begin{equation}
\chi^2 = \frac{1}{N} \sum\limits_{i=1}^N \frac{(\Delta X - m_X)^2}{\delta X^2}
\end{equation}
Here $\Delta X$ are the observed amplitudes, $\delta X$ their errors (both from Table \ref{amps}), $m_X$ the 
amplitudes from the simulations and $N$ is the number of filters. A good fit will result in $\chi^2 <1$. 

In Fig. \ref{f6}, upper row, we show the results for the CAFOS data; all parameter combinations with $\chi^2 <1$ are 
plotted with crosses, the ones with $\chi^2<3$ with dots. In both types of simulations, the observed amplitudes are 
best-matched by cool spots with $f \sim 0.1$. Not surprisingly, the match is better when the spot is modeled with the
AMES-DUSTY spectrum, as cool spots are expected to have a spectrum resembling a cool photosphere. Hot spots, on the
other hand, do not provide a good match. This result becomes stronger when we consider that the CAFOS amplitudes
are likely to be somewhat smaller, as the lightcurves are affected by atmospheric effects (Sect. \ref{lc}). 

The variations on timescales of 1-3 weeks are constrained by the BUSCA lightcurve and the Sloan photometry.
Here the amplitudes tend to be somewhat larger than on the 4\,d timescale covered by CAFOS. Moreover, the
amplitude ratio between the R- and I-band in the Sloan data is a factor of 3 larger than in the CAFOS 
lightcurves. The Sloan photometry indicates a steep increase of the amplitudes towards shorter wavelengths, 
which is typical for hot spots, as already discussed in \citet{2010MNRAS.408.1095S}. We ran the same spot 
simulations as before for the Sloan amplitudes (2nd section in Table \ref{amps}) and plot the results in 
Fig. \ref{f6} (lower left panel). This time only hot spots with $f<0.1$ provide a decent match. Not shown 
are the results for spots with AMES-DUSTY spectrum, because this series of simulations does not give any 
match with $\chi^2<3$. 

The best way to constrain the long-term optical variations on timescales of years is the comparison 
between our dataset and Sloan in the R- and I-band (3rd section in Table \ref{amps}). Here the R-band 
amplitude is by a factor of 4.5 larger than the I-band amplitude. Again, the simulations essentially 
rule out cool spots, but give a good match for a variety of parameter combinations for hot spots 
(Fig. \ref{f6}, lower right panel). As this test is based only on two bands, the results do not provide
particularly good constraints on the spot parameters. The simulation for the two Sloan epochs (see
above) with five filters, including the blue bands, probably gives a better idea of the hot spot 
properties.

We also explored the possibility of having hot and cool spots simultaneously on the surface of FU Tau A.
Using the Sloan amplitudes in five bands and adopting a filling factor of 5\% (for hot and cool spots, respectively) 
yields a good match with $\chi^2<1$ for $T_{\mathrm{hot}} = 3500-3600$\,K and $T_{\mathrm{cool}} = 2000-2900$\,K. 
For higher filling factors, $T_{\mathrm{hot}}$ decreases, and vice versa. Thus, the combination of hot and cool 
spots definitely fits the observed amplitudes.

We conclude that the periodic modulation on a timescale of 4\,d is best explained by cool spots co-rotating 
with the object. On longer timescales of weeks to years, however, the optical variability is dominated by hot 
spots. The cool spots are asymmetrically distributed and thus cause a periodic, rotational signal. The hot 
spots, on the other hand, are axisymmetric (e.g., limited to the polar regions) and stable in size/location 
over timescales of at least 4\,d, but vary on longer timescales. 

The cool spots are most likely caused by suppressed convection due to magnetic field lines penetrating through
the photosphere, as commonly observed for magnetically active stars, including the Sun. The hot spots could
be indicative of the same phenomenon: If most of the surface is covered by cool spots, the few remaining areas
of unspotted photosphere would be observed as hot spots. This would, however, require implausibly large filling 
factors of $>70$\%, because the hot spot solutions in the simulations give filling factors below 30\%. 

The more probable option is that the hot spots are shock fronts caused by gas accretion from the disk onto the 
object. This is also supported by the presence of H$\alpha$ emission in the spectra, which is likely to be 
dominated by accretion as well (Sect. \ref{spec}). The lack of variability in H$\alpha$ over timescales of 4\,d 
(EW varies by $<10-15$\%) is in line with the presence of stable accretion-related spots. If accretion is 
the origin of the hot spots, the long-term variability in FU Tau A indicates substantial changes in the accretion 
configuration (geometry or accretion rate) on timescales of years.

In addition to the optical variability, FU Tau also exhibits moderate changes of $\sim 0.2$\,mag at 
wavelengths of 3-8$\,\mu m$ on timescales of years (4th section in Table \ref{amps}), which are significantly larger
than the photometric error. The best explanation for the infrared variations is changes in the disk structure 
or temperature, which could be related to changes in the accretion flow.

\section{Discussion: The nature of FU Tau A}
\label{nature}

FU Tau A is anomalous in its observed properties, primarily in two aspects: 

1) In comparison with Taurus objects of similar spectral type and temperature, FU Tau A shows strong X-ray 
emission, although still consistent with the large scatter \citep[see Fig. 3 in][]{2010MNRAS.408.1095S}. 
Moreover, the X-ray emission is dominated by a soft radiation component, which may be explained by emission 
from an accretion shock. 

2) FU Tau A is overluminous in the HR diagram, with respect to the other known brown dwarfs of Taurus and 
to the theoretical 1\,Myr isochrone \citep[see Fig. 4 in][]{2010MNRAS.408.1095S}. This has been shown based 
on a luminosity calculated from the J-band magnitude \citep[$L_{\mathrm{bol}}/L_{\odot}=0.2$,][]{2009ApJ...691.1265L}. 
We re-determined the luminosity by comparing model spectra with the full set of optical and near-infrared 
photometry (SDSS ugriz, 2MASS JHKs) using VOSA \citep{2008A&A...492..277B}, and find 
$L_{\mathrm{bol}}/L_{\odot}=0.19\ldots0.21$, confirming the literature value. 

Since the overluminosity of FU Tau A is central for the following arguments, we verify this claim by plotting
FU Tau A and B in a magnitude vs. effective temperature diagram together with other late-type members
of Taurus. As a comparison
sample, we use the census by \citet{2010ApJS..186..259R}, which comprises spectroscopically confirmed
members of Taurus. Their sample of previously known members contains 215 objects, out of which 186 have a spectral 
type in the literature and 2MASS photometry. We limit ourselves to objects with spectral type later or equal M4, 
$A_V<5$\,mag. Since FU Tau A and B are a Class II source based on their SED and have low extinction 
\citep{2009ApJ...691.1265L}, we also exclude sources with the SED type 'Class I' or 'flat' and with 
$A_V>5$\,mag. This leaves 71 objects out of which 35 have a disk according to the Spitzer data and 20 
are detected in X-rays according to 
XEST\footnote{XMM-Newton Extended Survey of the Taurus molecular clouds} \citep{2007A&A...468..353G}.

\begin{figure}
\includegraphics[width=6.1cm,angle=-90]{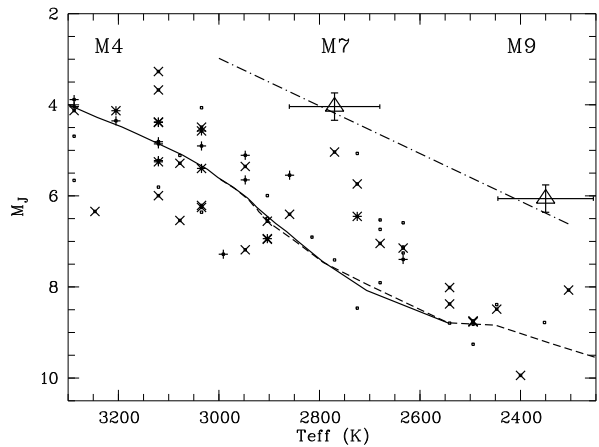}
\caption{Absolute J-band magnitude vs. effective temperatures for FU Tau A and B (triangles) and other confirmed
members of Taurus (dots, see text for selection criteria). Objects with disks are marked with crosses, those
with X-ray detection with plusses. We overplot the BCAH (solid lines) and DUSTY (dashed lines) ischrone for
an age of 1\,Myr \citep{1998A&A...337..403B,2001ApJ...556..357A}. The dashdotted line is a linear fit to the
DUSTY isochrone shifted to match the positions of FU Tau A and B.
\label{f7}}
\end{figure}

For these objects we determined $A_V$ from the $J-K$ colour:
\begin{equation}
A_V = [(J-K) - (J-K)_0] / 0.1844
\end{equation}
This is based on the extinction law by \citet{1989ApJ...345..245C} for $R_V = 4.0$. We use $(J-K)_0 = 1$ which is 
appropriate for this spectral type range, as outlined in \citet{2009ApJ...702..805S}. After correcting the J-band
magnitudes for extinction, we subtracted a distance modulus of 5.73\,mag for $d=140$\,pc, to yield the absolute J-band
brightness $M_J$. Spectral types were converted to effective temperatures using the relation by \citet{2008ApJ...689.1127M}.
For FU Tau A and the companion B we carried out the same procedure. For FU Tau A we adopted a spectral type of M7,
the average of our result (Sect. \ref{spec}) and the type given by \citet{2009ApJ...691.1265L}. Since no J-band 
magnitude is available for FU Tau B, we started with its K-band magnitude \citep{2010ApJ...720.1781L} and added 
1.3\,mag, which is the typical extinction corrected $J-K$ colour for Taurus members at spectral type $\sim$M9 or 
later. The resulting brightness-temperature diagram is shown in Fig. \ref{f7}.

Our goal was to minimise systematic effects that could influence a comparison between FU Tau A and the other
objects. The x-axis is affected by inconsistencies in spectral typing, which we estimate to be in the range of 
$\pm 0.5$ subtypes, as most of the objects have been classified using similar procedures and in the optical wavelength 
regime. On the y-axis, we estimate an uncertainty of $\pm 0.3$\,mag which mostly comes from the extinction estimate. 
One problem could be additional flux from disks in the K-band, which would cause us to overestimate extinction and 
brightness of objects with disks. In Fig. \ref{f7}, however, no systematic difference is seen between objects with 
disks and those without, indicating that this effect is negligible. 

Fig. \ref{f7} clearly demonstrates that FU Tau A is significantly brighter than all other similar objects in 
the Taurus star forming region with the same spectral type. The difference is $\sim 1$\,mag compared with the
brightest other objects at this spectral type or $\sim 2$\,mag compared with the average brightness at this 
spectral type. As can be seen in the figure, the same offset is observed for FU Tau B, i.e. it cannot be
explained by an effect that would only apply to FU Tau A, for example, an undetected close companion. One could
explain the offset by lowering the distance of FU Tau to 90\,pc or less, but this would place it outside
the Taurus star forming region which makes it difficult to explain the young age of the system. In summary,
the overluminosity of FU Tau A as well as the overluminosity of its companion is confirmed by our analysis.

In \citet{2010MNRAS.408.1095S} we suggest three plausible interpretations to explain the 2 anomalies discussed
above: suppressed convection, accretion, and the evolutionary stage of the object. In the following 
we discuss these scenarios in light of the new analysis presented in this paper. 

\subsection{Suppressed convection}

Strong magnetic activity could suppress the convection on the stellar surface and produce cool spots, as it seems
to be the case in the primary of the eclipsing brown dwarf binary 2MASS J05352184-0546085 \citep{2007ApJ...664.1154S}. 
This would make FU Tau A appear cooler than it should be according to its mass,  i.e. it would in fact 
be a very low mass star and not a brown dwarf. Correcting for this effect would shift FU Tau A roughly horizontally 
in the HR diagram towards higher temperatures and closer to the bulk of datapoints in Fig. \ref{f7}. The same
explanation could theoretically apply to FU Tau B.

A higher mass would also help to explain the high X-ray luminosity \citep{2010MNRAS.408.1095S}, because of the 
well-known correlation between $L_X$ and $L_\mathrm{bol}$ \cite[e.g.][]{2007A&A...468..425T}. In Fig. \ref{f7} we 
mark all objects with X-ray detection with plusses. The fraction of X-ray detected objects is only 2/29 (7\%) for 
$T_\mathrm{eff}<2800$\,K. In this temperature regime, FU Tau A is clearly an exception. For higher temperatures 
(2800-3300\,K), the rate is 18/42 (42\%) and the X-ray luminosity of FU Tau A would not be anomalous. In contrast
to FU Tau A, the companion B is not detected in X-rays \citep{2010MNRAS.408.1095S}, which fits well into the
lack of X-ray detections for objects below 2800\,K.

The scenario of suppressed convection would require fast rotation, strong magnetic fields and/or 
magnetic spot coverage \citep{2007A&A...472L..17C,2009ApJ...700..387M}. The only way to definitely confirm it
would be a model-independent mass estimate, which is not available for FU Tau A. 

Indirectly, however, the new data supports the idea of suppressed convection. The presence of cool, magnetic spots 
is confirmed by the lightcurve analysis (Sect. \ref{disc}), although the lightcurves can only detect asymmetrically
distributed spots and cannot be used to fully characterise the spot distribution on the stellar surface. The rotation 
period of 4\,d is quite long when compared with typical periods for young brown dwarfs \citep[$<2$\,d,][]{2005A&A...429.1007S}, 
but fits much better with the typical range of periods for young very low mass stars \citep[2-5\,d,][]{2001ApJ...554L.197H}. 
Thus, suppressed convection remains a plausible option.

\subsection{Accretion}

Accretion from a circumstellar disk might cause excess luminosity in the optical and near-infrared and shift objects
vertically in Fig. \ref{f7}. As said above, accretion might also be the best explanation for the soft X-ray emission in 
FU Tau A. Accretion is clearly ongoing, as evidenced by the optical spectra (this paper, Stelzer et al. in prep.). The long-term 
variability in FU Tau A can only be explained by excess flux from hot spots and could indicate variable accretion 
(Sect. \ref{disc}). However, since the spot temperatures are not particularly high ($<5000$\,K) accretion is unlikely 
to cause a drastic over-estimate in the luminosity. For plausible spot parameters of $T_S = 4000$\,K and 
$f=0.1$ (see Fig. \ref{f6}) the excess flux at 1.0$\,\mu m$ would only be in the range of 30\%. It needs to be stressed, 
however, that the variability is caused only by the changes in the hot spots, i.e. higher excess luminosity from constant, 
axisymmetrically distributed spots is possible. 

The interpretion of the high luminosity in terms of accretion would imply that the actual $L_{\mathrm{bol}}$ is 
much lower, leading to an unusually high fractional X-ray luminosity. If the luminosity is in fact one
order of magnitude lower than given by \citet{2009ApJ...691.1265L}, we would get $\log{(L_X/L_{\mathrm{bol}})} = -2.2$
\citep{2010MNRAS.408.1095S}, much more than the median for Taurus brown dwarfs \citep[-4.0,][]{2007A&A...468..391G},
which would thus create a new anomaly. Also, some other very low mass objects in Taurus share the unusual characteristics 
of FU Tau A (position in Fig. \ref{f7} above the isochrone and strong X-ray emission), albeit not as extreme as our target, 
and some of them are not accreting. For these reasons, accretion is unlikely to be the only reason for the anomalies of 
FU Tau A.

Similar to FU Tau A, the companion FU Tau B shows strong H$\alpha$ emission and mid-infrared excess indicative of
the presence of a disk \citep{2009ApJ...691.1265L}. Thus, accretion could also affect its position in Fig. \ref{f7}.

\subsection{Evolutionary stage}

The overluminosity of FU Tau A and B could also be caused by extreme youth.  If accretion occurs mostly in bursts 
('episodic accretion'), objects  which are technically 'coeval', i.e. have started their protostellar collapse at the 
same time, could be in a different stage of their accretion history and spread out in the HR diagram 
\citep{2009ApJ...702L..27B}. Overluminous objects would be younger in terms of their accretion evolution. 

In Fig. \ref{f7} we overplot the BCAH and DUSTY 1\,Myr isochrones \citep{1998A&A...337..403B,2001ApJ...556..357A}.
The comparison with the Taurus members shows that most of them fall within $\pm 1$\,mag of this isochrone, indicating
an average age of 1\,Myr. The dash-dotted line is a linear fit to the DUSTY isochrones, shifted by 3\,mag towards
brighter absolute magnitudes. The locations of FU Tau A and B are well-approximated by this line, i.e. the two
objects sit about 3\,mag above the isochrone. The Baraffe et al. models do not include the formation process and 
accretion history \citep{2002A&A...382..563B}, i.e. the comparison with the observations is limited. It does 
show, however, that FU Tau A and B have roughly equal distance from the isochrone on the y-axis, i.e. in terms 
of these models they can be considered to be coeval. 

If extreme youth is the reason for the unusual brightness of the two FU Tau components, it seems surprising that no 
other typical signs of the earliest evolutionary stages of T Tauri stars are seen. The objects have low extinction 
and their SEDs are classified as Class II, i.e. they are not embedded in a thick protostellar envelope. Also, the 
current accretion rate of FU Tau A is relatively low and comparable to the other Class II objects of similar spectral 
type in Taurus \citep[$10^{-9}-10^{-10}\,M_{\odot}\,$yr$^{-1}$,][]{2010MNRAS.408.1095S}. In addition, Class 0/I protostars
appear to show lower or similar X-ray luminosities than Class II sources \citep{2008ApJ...677..401P}, e.g. extreme youth 
cannot explain the strong X-ray emission of FU Tau A. Thus, an early evolutionary stage is not an entirely satisfactory 
explanation.

\subsection{Conclusions}

In summary, FU Tau A is affected by accretion plus magnetic activity and possibly evolutionary stage as well. Although the 
observed properties of FU Tau A are anomalous, these three factors are not limited to this particular object. All young 
populations show substantial spread in the HR diagram and in the other relevant properties (e.g., X-ray luminosities, 
H$\alpha$ luminosities, rotation periods). Evidence for accretion and strong magnetic activity is commonly seen in 
classical T Tauri stars and brown dwarfs. FU Tau A (and its companion FU Tau B) happens to be a case which sticks out 
in some of its observed properties, but there is no reason to believe that it is unusual in its physical characteristics. 

This could have important consequences for our understanding of the Initial Mass Function in the low-mass regime. Even with 
the wealth of photometric and spectroscopic data available for FU Tau A, it is still not possible to estimate its mass 
reliably. Depending on the relative magnitude of the effects of suppressed convection and accretion, the mass could
be anywhere between 0.05 and 0.2$\,M_{\odot}$, a factor of 4 uncertainty. In addition, the unknown evolutionary stage 
and accretion history means that we cannot trust the isochrones for mass estimates. Thus, barring a more complete 
understanding of magnetic activity and its effect on the observable properties as well as protostellar evolution 
it does not seem feasible to derive a mass function for very young very low mass stars and brown dwarfs.

\section*{Acknowledgments}
We thank the anonymous referee for a constructive report that helped to improve the paper.
This publication is based on observations collected at the Centro Astron\'omico Hispano Alem\'an 
(CAHA) at Calar Alto, operated jointly by the Max-Planck Institut f{\"u}r Astronomie and the Instituto 
de Astrof\'isica de Andaluc\'ia (CSIC). Part of this work was funded by the Science Foundation Ireland 
through grant no. 10/RFP/AST2780 to AS, and the Spanish grants AyA2010-21161-C02-02, CSD2006-00070,
PRICIT-S2009/ESP-1496. In this paper we make use of VOSA, developed under the Spanish Virtual Observatory 
project supported from the Spanish MICINN through grant AyA2008-02156.

\section*{Note added in proof}

Regarding the discussion in Sect. \ref{calphot}, it was brought to our 
attention that the SDSS photometry for FU Tau A and B was published in the 
data release of the 'Low Galactic Latitude Fields', which are not part of 
the standard SDSS data releases  \citep{2004AJ....128.2577F}. The correct epoch 
for the SDSS datapoints for FU Tau is Dec 31 (Luhman, private communication). 
 
\newcommand\aj{AJ} 
\newcommand\actaa{AcA} 
\newcommand\araa{ARA\&A} 
\newcommand\apj{ApJ} 
\newcommand\apjl{ApJ} 
\newcommand\apjs{ApJS} 
\newcommand\aap{A\&A} 
\newcommand\aapr{A\&A~Rev.} 
\newcommand\aaps{A\&AS} 
\newcommand\mnras{MNRAS} 
\newcommand\pasa{PASA} 
\newcommand\pasp{PASP} 
\newcommand\pasj{PASJ} 
\newcommand\solphys{Sol.~Phys.} 
\newcommand\nat{Nature} 
\newcommand\bain{Bulletin of the Astronomical Institutes of the Netherlands}
\newcommand\memsai{Mem. Soc. Astron. Ital.}

\bibliographystyle{mn2e}
\bibliography{aleksbib}

\label{lastpage}

\end{document}